\documentclass[12pt, letterpaper]{article}
\usepackage[top=1in, bottom=1.5in, left=1in, right=1in]{geometry}
\usepackage[utf8]{inputenc}
\usepackage{booktabs}
\usepackage{hyperref}

\usepackage{amsmath,amssymb}
\usepackage[numbers,sort&compress]{natbib}
\usepackage{xcolor}
\usepackage{graphicx}
\usepackage{float}

\newcommand{\wfirst}{{\em WFIRST }}
\newcommand{\wfirstns}{{\em WFIRST}}
\newcommand{\lsst}{{\em LSST }}
\newcommand{\lsstns}{{\em LSST}}

\title{Unique Science from a Coordinated LSST-WFIRST Survey of the Galactic Bulge}
\author{R.A.~Street, M.B.~Lund, M.~Donachie, S. Khakpash, N.~Golovich, \\
M.~Penny, D.~Bennett, W.A.~Dawson, J.~Pepper, M.~Rabus, \\
P.~Szkody, W.I.~Clarkson, R.~Di Stefano, N.~Rattenbury, \\
M.P.G.~Hundertmark, Y.~Tsapras, S.~Ridgway, K.~Stassun, V.~Bozza, \\
A.~Bhattacharya, S.~Calchi Novati,Y.~Shvartzvald, \\with the support of the \\LSST Transient and Variable Stars Collaboration.}
\date{Nov 2018}

\begin{document}

\maketitle

\begin{abstract}
NASA's \wfirst mission will perform a wide-field, NIR survey of the Galactic Bulge to search for exoplanets via the microlensing techniques.  As the mission is due to launch in the mid-2020s, around half-way through the LSST Main Survey, we have a unique opportunity to explore synergistic science from two landmark programs.  \lsst can survey the entire footprint of the \wfirst microlensing survey in a single Deep Drilling Field.  Here we explore the great scientific potential of this proposal and recommend the most effective observing strategies.  
\end{abstract}

\section{White Paper Information}

\noindent{\bf Science Category: } Exploring the transient optical sky, mapping the Milky Way \\
\noindent{\bf Survey Type Category: } Deep Drilling Field\\
\noindent{\bf Observing Strategy Category: } Integrated program with science that hinges on the combination of pointing and detailed observing strategy\\

\noindent{\bf Author's contact information:}\\

R.A.~Street, Las Cumbres Observatory, rstreet@lco.global

M.B.~Lund, Vanderbilt University, michael.b.lund@vanderbilt.edu

M.~Donachie, University of Auckland, m.donachie@auckland.ac.nz

S. Khakpash, Lehigh University, somayeh.khakpash@gmail.com 

N.~Golovich, Lawrence Livermore National Laboratory, golovich1@llnl.gov,

M.~Penny, Ohio State University, penny@astronomy.ohio-state.edu

D.~Bennett, NASA Goddard Space Flight Center, david.p.bennett@nasa.gov

W.A.~Dawson, Lawrence Livermore National Laboratory, will@dawsonresearch.com

J.~Pepper, Lehigh University, joshua.pepper@lehigh.edu

M.~Rabus, Pontificia Universidad Cat\'olica de Chile, markus.rabus@gmail.com

P.~Szkody, University of Washington, szkody@astro.washington.edu 

W.I.~Clarkson, University of Michigan, wiclarks@umich.edu

R.~Di Stefano, CfA, Harvard, rdistefano@cfa.harvard.edu

N.~Rattenbury, University of Auckland, n.rattenbury@auckland.ac.nz

M.P.G.~Hundertmark, Universit\"{a}t Heidelberg, 

markus.hundertmark@uni-heidelberg.de

Y.~Tsapras, Universit\"{a}t Heidelberg, ytsapras@ari.uni-heidelberg.de

S.~Ridgway, NOAO, ridgway@noao.edu

K.~Stassun, University of Vanderbilt, keivan.stassun@vanderbilt.edu

V.~Bozza, Università di Salerno, valboz@sa.infn.it

A.~Bhattacharya, University of Maryland, abhatta5@umd.edu

S.~Calchi Novati, Caltech, snovati@ipac.caltech.edu

Y.~Shvartzvald, Jet Propulson Laboratory, yossishv@gmail.com
\vspace{0.5cm}

Street, Penny, Bennett, Stassun, Bozza, Bhattacharya, Calchi Novati and Shvartzvald are representatives of the WFIRST Microlensing Science Investigation Team (Bennett is deputy team lead).  
\clearpage

\section{Scientific Motivation}
\label{sec:science_case}

\begin{figure}[H]
\begin{centering}
\begin{tabular}{cc}
\includegraphics[width=0.525\textwidth]{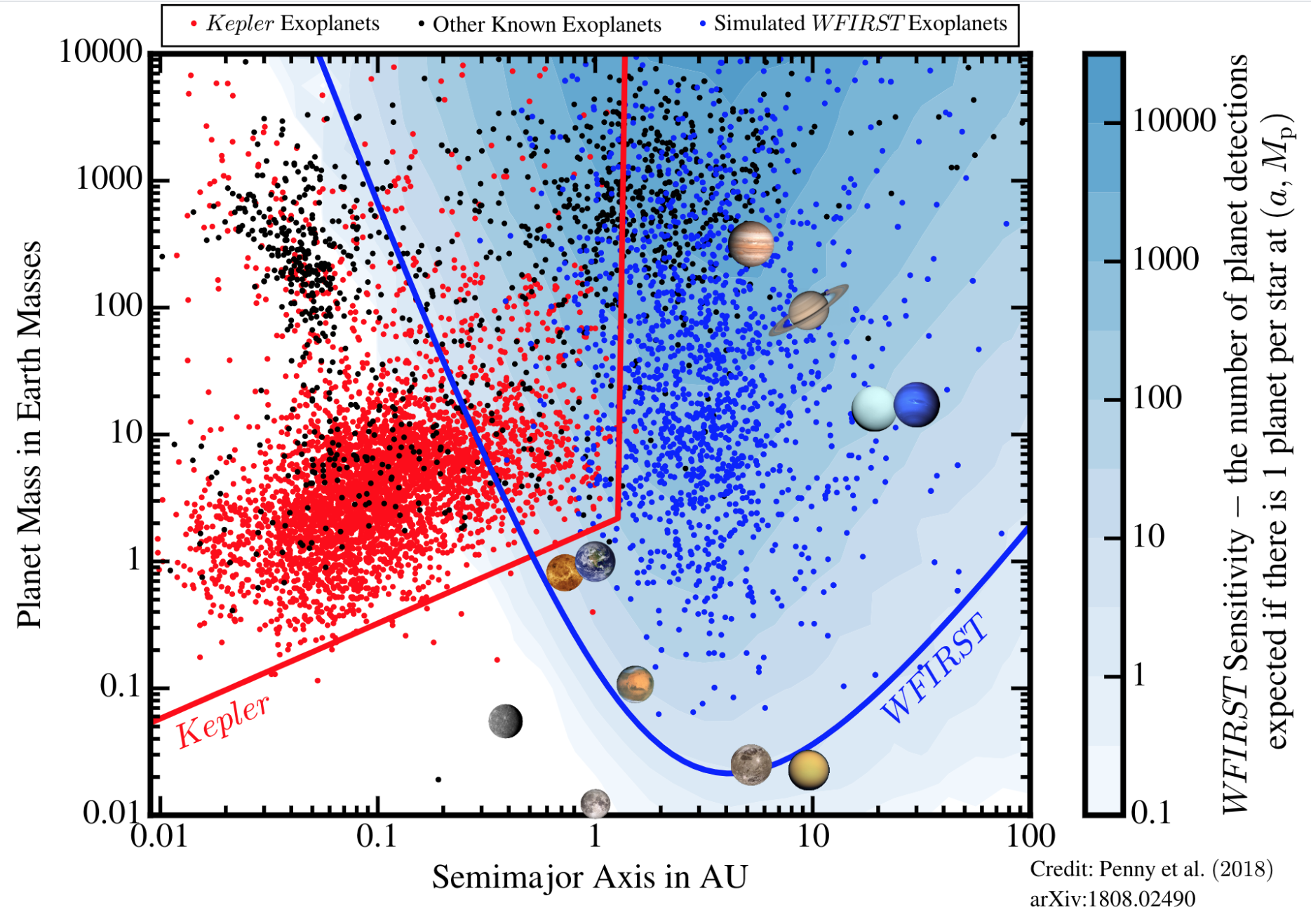}&
\includegraphics[width=0.475\textwidth]{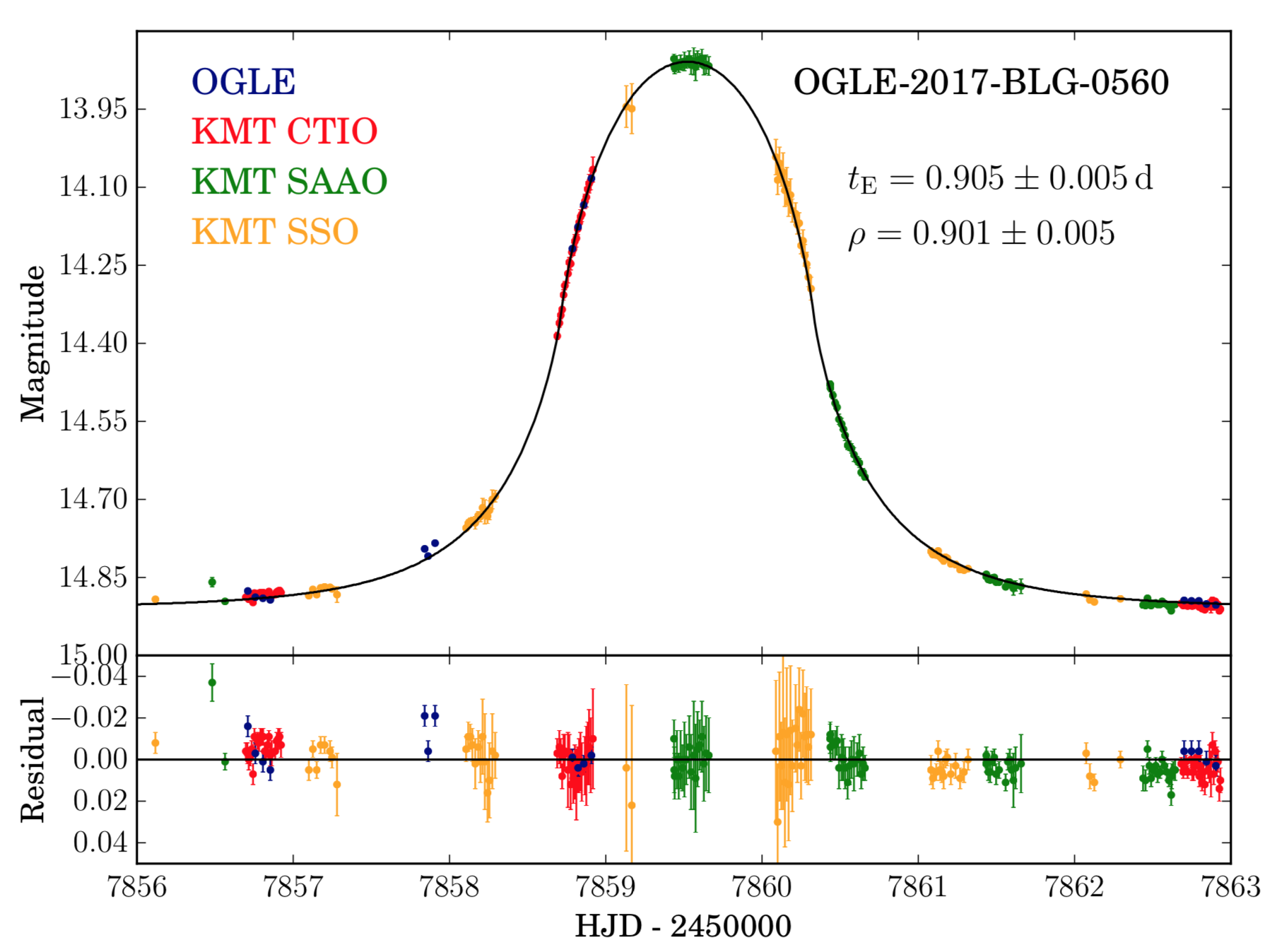}
\end{tabular}
\caption{(Left) Distribution of masses and orbital separations of known exoplanets, color/symbol-coded by detection method.  The shaded region indicates the parameter space where microlensing is sensitive.\label{fig:Mvsa} (Right) Lightcurve of FFP candidate event OGLE-2017-BLG-0560 from \cite{Mroz2018}.\label{fig:FFPlc}}
\end{centering}
\end{figure}

\begin{figure}[H]
\begin{centering}
\begin{tabular}{cc}
\includegraphics[width=0.45\textwidth]{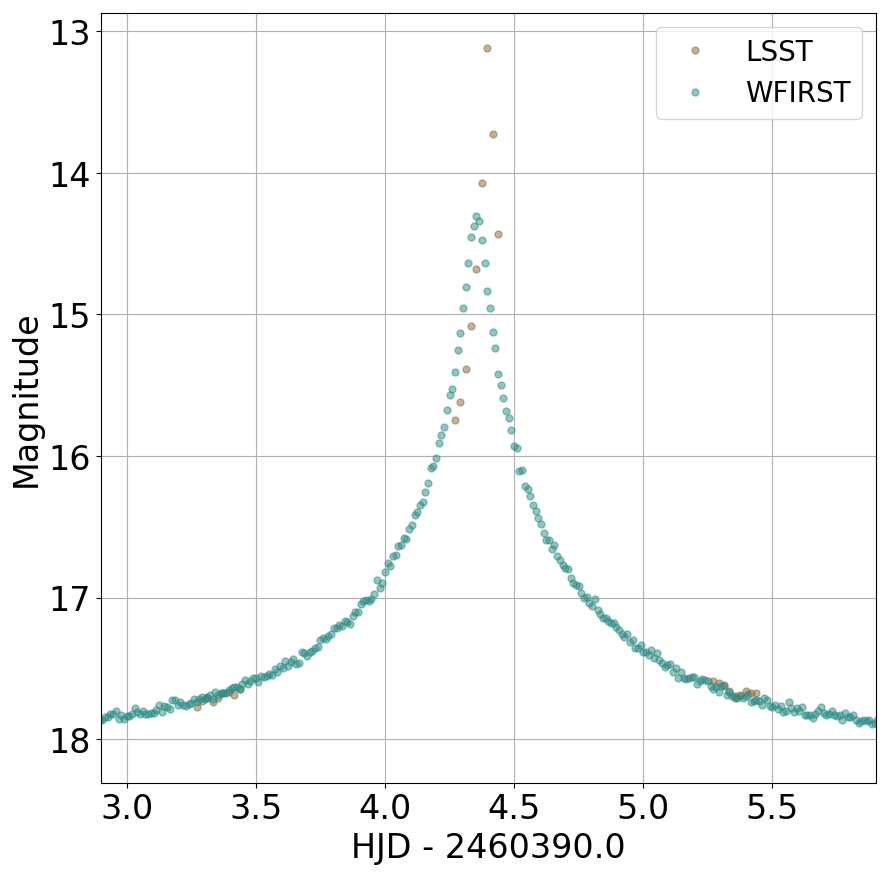}
\includegraphics[width=0.45\textwidth]{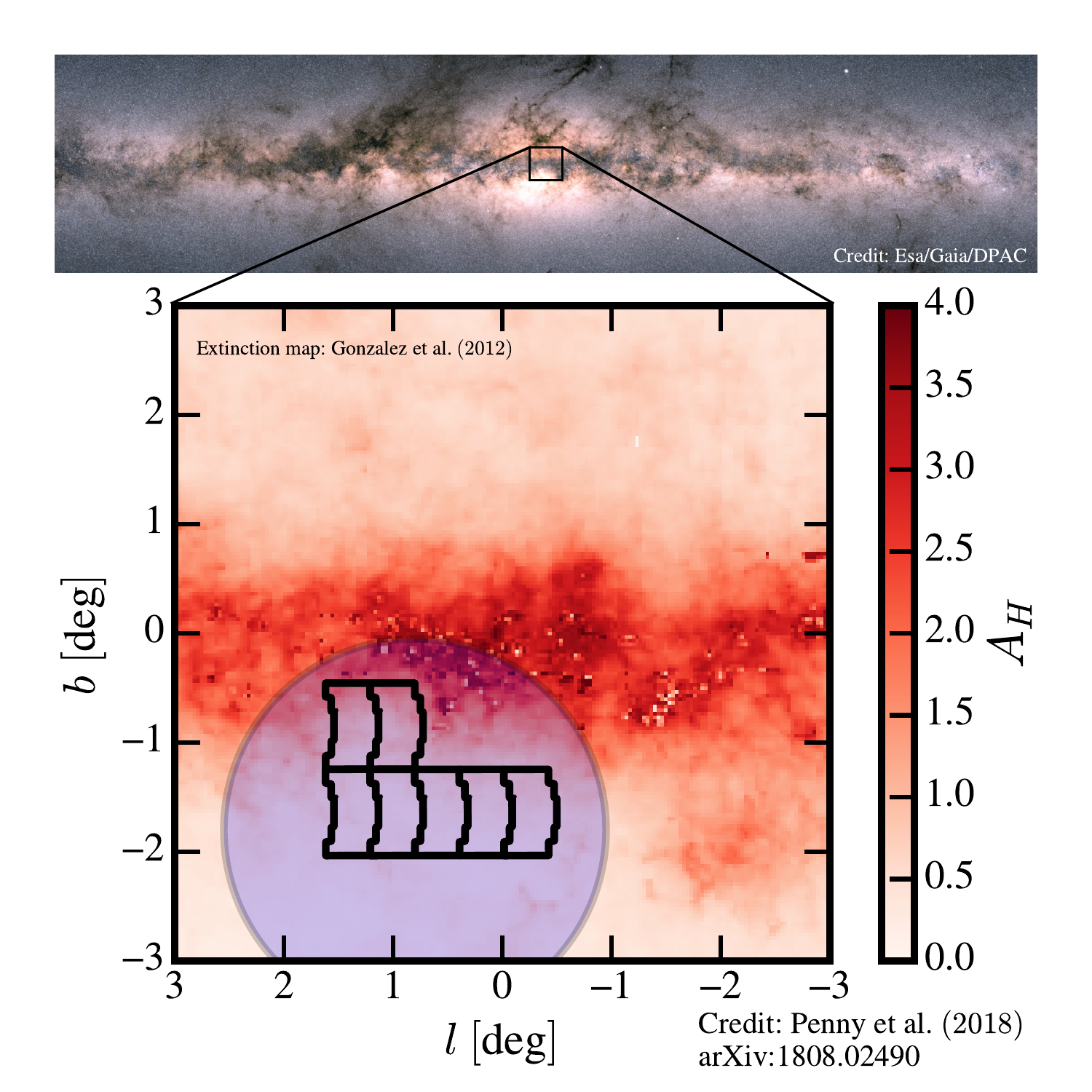}
\end{tabular}
\caption{(Left) Simulated simultaneous lightcurves of a free-floating-planet microlensing event. \label{fig:FFP} (Right) Comparision of the WFIRST Bulge Survey field (black mosaic outline, 1.53$\times$1.5$^{\circ}$) with the field of view of an LSST Deep Drilling Field (blue circle, 3.5$^{\circ}$ diameter).\label{fig:fov}}
\end{centering}
\end{figure}

\noindent{\bf Exoplanets - microlensing: }One of our most powerful tools for understanding planetary formation is to compare the actual planet population with that predicted by simulations, but there remain important gaps in our planet census (Fig.~\ref{fig:Mvsa}).  Low-mass planets in orbits between $\sim$1--10 AU are of particular interest, because the core accretion mechanism predicts a population of icy bodies (e.g. \cite{IdaLin2013}) in this region. Evolutionary models further predict that gravitational interactions between migrating protoplanets should result in some being ejected from their systems \cite{Mustill2015, Chatterjee2008}.  However, this parameter space coincides with a gap in the sensitivity of the planet-hunting techniques used to date, leading to it being sparsely sampled.  Microlensing offers a way to test both of these predictions, being capable of detecting planets down to $\sim$0.1\,$M_{Earth}$ at orbital separations of $\sim$1--10\,AU.   Microlensing occurs when a massive body (lens) passes directly between an observer and a background source \cite{Tsapras2018}.  The gravity of the lens deflects light from the source, causing it to brighten and fade as the objects move, with relative velocity $\overline{\nu}$.  The crossing timescale for the event, $t_{E}$, is proportional to the mass of the lens.  Uniquely, microlensing requires no light from the {\it lens}, enabling it to detect objects otherwise too dark to measure, including free-floating planets [FFPs].  The effectiveness of this technique has now been demonstrated by the discovery of three candidate FFP events \cite{Mroz2017, Mroz2018} (Fig~\ref{fig:FFPlc}).  A survey of the Galactic Bulge region where the rate of microlensing events is highest is  one of the main goals of NASA's \wfirst Mission \cite{Spergel2015}.  The spacecraft will discover $\sim$1400 bound planets and will provide a dataset ideal for detecting FFPs \citep{Penny2018}.  The physical properties of bound planets should be constrained by direct measurement of the light from the lensing system, but this technique cannot be applied for FFPs.  Microlensing models suffer from a number of degeneracies, and the physical properties of the lens are extremely hard to measure without an additional constraint on the event parallax.  For long timescale events ($>$30\,d) this can be derived from a single lightcurve thanks to the orbital motion of the observer, but \wfirst data alone cannot measure the parallax for short timescale ($t_{E}$\,$\leqslant$30\,d) events.  Thanks to the $\sim$0.01\,AU separation between \lsst and \wfirst (at L2), the  observatories will measure different magnifications and times of maximum (see Fig.~\ref{fig:FFP}), enabling us to derive the physical and dynamical properties of short timescale events.  {\bf \lsst will substantially improve constraints on the lens properties, particularly the FFP mass function, distances and kinematics, by simultaneously observing a Deep Drilling Field covering the \wfirst Bulge survey region.} \lsst will also complete event lightcurves that remain partially sampled by \wfirstns.  \wfirst will observe $i\sim$ 19--25\,mag stars in the Bulge for a total of $\sim$432\,d spread over 6 `seasons'.  But the spacecraft can only monitor the field for $\sim$72\,d at a time due to pointing constraints.  Since microlensing events can peak at any time, and have durations $t_{E}\sim $1--100\,d, many \wfirst lightcurves will be incompletely sampled.  This will make it difficult to measure the parallax (and hence physical properties) for even long timescale events, and raises the probability that anomalous features (and lens companions) will be missed in the inter-season gaps.  \lsst is ideally suited for this science: it can monitor {\it all} events down to $i\sim $23\,mag (see field-of-view, Fig.~\ref{fig:fov}); no other survey facility provides the combination of etendue {\it and} baseline in all filters (DECam is limited by detector sensitivity in redder bandpasses; Pan-STARRS-1 has restricted visibility during simultaneous windows).  \lsst and \wfirst combined will deliver complementary photometry in both optical and NIR for millions of stars, which will in addition facilitate a wide range of stellar astrophysics (see white papers by Dell'Ora et al. and Lund et al.).  

\noindent{\bf Exoplanets - transits: }The high cadence, high precision photometry for millions of Bulge stars produced by \wfirst will also be an ideal data in which to discover $\sim$10$^{5}$ planetary transits \cite{Montet2017}.  Owing to the (relatively faint) candidate magnitudes, traditional radial velocity follow-up will not be possible in most cases, so optical photometry from \lsst will greatly help to rule out astrophysical false positives. 

\noindent{\bf Variability in Ultra-cool dwarfs (UCDs): } Photometric monitoring of brown dwarfs has shown that some of these objects show a time-dependent variability around their rotation periods of $\sim$8\,hrs \citep{Artigau2018}.  While cloudy atmospheres could explain the observed variations \cite{marley:2010}, many problems can only partially be explained, as e.g. resurgence of FeH absorption, so a number of alternatives have been proposed.  These include non-uniform temperature profiles \citep{robinson:2014}, thermochemical instabilities \citet{tremblin:2016} and lightning and auroral activities \citet{Helling:2013, Bailey:Lightning, Hodosan:Lightning}.  However, the sample of these intrinsically faint objects with long baseline data is too small to draw meaningful conclusions.  Combining simultaneous high cadence optical and NIR data probes different layers and features of their atmospheres.  This DDF $+$ \wfirst will deliver extended-baseline data for tens of new objects \citep{Reyle2010}.  

\noindent{\bf Bulge Globular clusters (GCs): } GCs are among the oldest stellar populations in the Galaxy and are rich in RR Lyrae, which are commonly used as standard candles to measure distances. Their ages, metallicities and distances can be used to determine the dynamical and physical conditions at the early stages of formation of the Galaxy, as well as stellar structure and evolution \citep{krauss:2003,roediger:2014}. But few of the 84 Bulge GCs \citep{minniti:2017} have been examined \citep{bica:2016,tsapras:2017} because the high reddening and differential extinction by foreground dust limits the accuracy of existing surveys. In contrast, \lsst will deliver high-resolution, deep multi-band photometry for $>$9 clusters within this DDF, and provide reliable estimates of the extinction as well as accurate color-magnitude diagrams, distances and metallicity measurements for the inner parts of the Galaxy \citep{bobylev:2017}. These results can then be compared with similar studies of halo clusters and used to calibrate models of Galactic formation and evolution \citep{binney:2017}.

\noindent{\bf Cataclysmic Variables (CVs): } CVs, close binaries with a white dwarf acccreting from a late main-sequence star or brown dwarf, represent the most common end-product of binary star evolution, and their frequency of occurrence places a constraint on stellar evolution models. However, an accurate estimate of the population has yet to be completed in the Plane due to their faintness and crowding.  High cadence data allows their orbital periods to be measured, testing the predictions of $P\sim $80\,min \cite{GoliaschNelson2015}, while longer-baseline data will detect dwarf novae in outbursts.  Their blue colors mean that \lsst data will complement the red/NIR data from \wfirstns, while both surveys will push $\sim$3\,mag fainter than the OGLE survey \cite{Mroz2015}, which indicated at least 264 dwarf novae brighter than $\sim$16$<$I$<$20\,mag within this field.  

\pagebreak

\vspace{.6in}

\section{Technical Description}
\begin{footnotesize}
{\it Describe your survey strategy modifications or proposed observations. Please comment on each observing constraint
below, including the technical motivation behind any constraints. Where relevant, indicate
if the constraint applies to all requested observations or a specific subset. Please note which 
constraints are not relevant or important for your science goals.}
\end{footnotesize}

\subsection{High-level description}
\begin{footnotesize}
{\it Describe or illustrate your ideal sequence of observations.}
\end{footnotesize}

We propose that LSST observe a Deep Drilling Field (DDF) that includes the WFIRST Bulge survey footprint.  As the \wfirst field of view is 0.281\,deg$^{2}$, \lsst will be able to survey the entire \wfirst  footprint with a single pointing.  We therefore request that \lsst locate a DDF on this region, timed to complement \wfirst observations:
\begin{itemize}
\itemsep0em 
\item Observations at 30min cadence during periods of simultaneous viewing with \wfirstns, to measure event parallax as well as stellar, brown dwarf and planetary variability,
\item Low (~3\,d) cadence observations during \wfirst inter-season gaps to complete event coverage and obtain long-baseline coverage of periodic variability,
\item Multi-wavelength observations to constrain stellar properties and variability, complementing \wfirstns's NIR passbands. 
\end{itemize}

For reference, the \wfirst Bulge survey will obtain continuous imaging of $\sim$7\,fields in the W149 (927--2000\,nm) with a 15\,min cadence for 6 survey periods of 72\,d duration.  These periods will be distributed through the spacecraft's primary mission (launch: 2025), and centered around the equinoxes due to pointing constraints.  Lower cadence imaging in filters Z087 (760--977\,nm), R062 (480--760\,nm) and F184 (1683--2000\,nm) will also be obtained at lower cadence, between 3--6\,hrs.  \\

\noindent {\bf Observing Constraints:}\\
\begin{enumerate}
    \item {\bf Visibility:} The periods when \wfirst can observe the Bulge are constrained by the spacecraft's orientation requirements relative to the Sun, and are limited to a window $\pm$36 days from the time when the direction of the Sun is 90 degrees from the \wfirst pointing direction.  In practice this means it will conduct 6 survey `seasons' between $\sim$Feb 12-- Apr 24 and $\sim$Aug 19 -- Oct 29, between $\sim$2026 to 2030. LSST observations of the same field will be constrained by the fields annual visibility from Earth, so simultaneous observations will only be possible betwen $\sim$late January and mid-October.  {\bf For much of these windows the field will only be observable for 1 -- few hours at the start and end of each night, limiting the time required for the DDF, and making it possible to schedule these observations around ongoing WFD and other surveys.}
    \item {\bf Bandpasses:} This DDF is a region of high optical extinction, making $u$-band observations difficult.  In order to minimize the time requirements for this DDF, we propose observations in a restricted filterset: $griz$. 
    \item {\bf Limiting magnitude:} The primary goal of this DDF is time coverage rather than to reach fainter limiting magnitudes.  We expect our targets to achieve adequate signal-to-noise in the 1$\times$30\,s exposure per visit.
    \item {\bf Moon: } The Moon passes close to the Galactic Bulge several times a year, which will necessarily interrupt observations of this field for a few days around the date of closest passage.  
\end{enumerate}

\noindent {\bf Proposed observing strategy:}\\
During periods of simultaneous observations with \wfirstns, we propose that LSST should obtain a single, 30\,s exposure of a DDF overlapping the \wfirst footprint once every 30\,min with an ``alternating-$i$'' sequence.  That is, every {\it other} visit to the field would use the $i$ filter, while the filter for the visit immediately following an $i$ exposure would rotate through the set $g,r,z$, i.e. $i - g - i - r - i - z - i - g - i - r$...  This results in cadences of 1\,hr in $i$-band and 3\,hrs for all other filters.  

This will effectively double the cadence in multiple colors, relative to \wfirst observations alone, providing better lightcurve sampling which will be useful in constraining the microlensing source star colors, but is particularly important for brown dwarf atmosphere studies. 

During the inter-season gaps and for rest of the Bulge visibility window from Earth (April-October) in 2026-2030, we propose that LSST image the \wfirst DDF footprint once a day, following the same ``alternating-$i$'' strategy.  However we note that if the WFD survey were extended to the Galactic Bulge, this requirement may be met automatically, and would not need to be included here. 

We note that a 15\,min cadence during the simultaneous periods would be optimal for our science because it would provide better constraints on the times of event peaks and caustic crossings, and would be
key to characterizing the very short durations FFP microlensing signals.  Based on the typical projected relative velocity of a Bulge event ($\sim$500\,km s$^{-1}$), the time delay between \wfirst and \lsst observing these events is expected to be of the order of 50\,min, so the higher sampling is likely to be beneficial.  Mindful of the additional time required by this cadence and the pressure on other programs we request the minimum viable sampling of every 30\,min, but raise it here for consideration.  We will conduct further investigations to quantify the improvement of using higher sampling.  

\vspace{.3in}

\subsection{Footprint -- pointings, regions and/or constraints}
\begin{footnotesize}{\it Describe the specific pointings or general region (RA/Dec, Galactic longitude/latitude or 
Ecliptic longitude/latitude) for the observations. Please describe any additional requirements, especially if there
are no specific constraints on the pointings (e.g. stellar density, galactic dust extinction).}
\end{footnotesize}\\

While the exact pointings and number of fields of the \wfirst Bulge survey has yet to be finalized, it will consist of a limited ($\sim$7) number of fields concentrated on the region of highest microlensing rate in the Galactic Bulge.  This has been determined from previous ground-based surveys to be {\bf centered around RA,Dec=(17:57:00, -29:13:00)}, and although \wfirst's NIR instrument can in principle survey fields of greater stellar density at lower galactic latitude, in practise microlensing events found here are expected to be more difficult to characterize, so most if not all of the \wfirst survey will concentrate in the region accessible to optical bandpasses.  As the \wfirst field of view is 0.281\,deg$^{2}$, \lsst will be able to encompass the entire \wfirst footprint with a single pointing.  This may be further optimized by offsetting the LSST DDF such that the \wfirst region is non-centered.  This avoids regions of high optical extinction and maximizes the number of stars surveyed, resulting in a corresponding increase yield in Galactic Plane science.  

There is no scientific advantage to dithering this pointing, since this would negatively impact the cadence for stars at the edge of the field of view.

\subsection{Image quality}
\begin{footnotesize}{\it Constraints on the image quality (seeing).}\end{footnotesize}\\

Owing to the high degree of stellar crowding in Bulge fields, photometric measurements will become significantly less precise during periods of poor seeing.  We therefore request observations in conditions of $<$2\,arcsec.  

\subsection{Individual image depth and/or sky brightness}
\begin{footnotesize}{\it Constraints on the sky brightness in each image and/or individual image depth for point sources.
Please differentiate between motivation for a desired sky brightness or individual image depth (as calculated for point sources). Please provide sky brightness or image depth constraints per filter.}
\end{footnotesize}\\
As the primary goal of this proposal is time-coverage rather than depth, we propose to observe in all sky brightness conditions, with the exception of the occasions when the Moon passes close enough to the Bulge to endanger the telescope.  In these instances, LSST's standard lunar-avoidance algorithm will be sufficient.   The limiting magnitude expected per bandpass for a single 30\,s exposure is estimated below, taking the expected extinction into account. 

\begin{table}[h!]
\centering
\begin{tabular}{lllll}
\hline
             & Sky brightness & Limiting magnitude \\
\hline
SDSS-$g$     &  All           &  23.0\\
SDSS-$r$     &  All           &  23.3\\
SDSS-$i$     &  All           &  22.9\\
SDSS-$z$     &  All           &  22.3\\
\hline
\end{tabular}
\end{table}

\subsection{Co-added image depth and/or total number of visits}
\begin{footnotesize}{\it  Constraints on the total co-added depth and/or total number of visits.
Please differentiate between motivations for a given co-added depth and total number of visits. 
Please provide desired co-added depth and/or total number of visits per filter, if relevant.}
\end{footnotesize}\\
Our science cases do not require the co-addition of images.  
If all possible visits are executed according to the requested cadence, we expect to make 746 visits in the spring simultaneous viewing period and 752 visits in the fall period.  117 visits will be made during the inter-season gap, making a total of 1615 visits per year between 2026-2030, assuming \wfirst observes the Bulge in both spring and fall.  

\subsection{Number of visits within a night}
\begin{footnotesize}{\it Constraints on the number of exposures (or visits) in a night, especially if considering sequences of visits.  }
\end{footnotesize}\\
During LSST/\wfirst periods of simultaneous visibility, we request as many visits as possible during time when the field is visible from \lsst.  The number of visits within a given night varies as it rises or sets from the Earth, from 3 visits/night at the start of February and mid-October to 13 visits/night in late spring/early fall.  

During the inter-season gap, the requested cadence is once per day.  

\subsection{Distribution of visits over time}
\begin{footnotesize}{\it Constraints on the timing of visits --- within a night, between nights, between seasons or
between years (which could be relevant for rolling cadence choices in the WideFastDeep. 
Please describe optimum visit timing as well as acceptable limits on visit timing, and options in
case of missed visits (due to weather, etc.). If this timing should include particular sequences
of filters, please describe.}
\end{footnotesize}

The timing of this DDF will necessarily have to coordinate with the launch and survey operations of the \wfirst Mission.  At time of writing, the mission is expected to launch in 2025 and begin survey operations in 2026.  We request that {\it low-cadence} DDF observations begin the year {\it prior} to the \wfirst survey starting, as this will provide a valuable catalog of variables within the field which {\it both} projects will use to optimize their selection of microlensing events.  

The exact periods when \wfirst will observe the Bulge are not yet finalized, but are expected to occur at 6-12\,month intervals from 2026 onwards.  We therefore request that \lsst DDF simultaneous observations be timed to coincide with \wfirst.  

Inter-season observations {\it between} \wfirst survey periods are most valuable immediately following or preceeding a \wfirst survey period.  

It should be noted that, although the Bulge visibility from the ground is very small at the very start and very end of the year, this is the most favourable period for the measurement of the microlensing parallax.

\subsection{Filter choice}
\begin{footnotesize}
{\it Please describe any filter constraints not included above.}
\end{footnotesize}\\

As described above, we request observations in $griz$, neglecting $u$ due to the high extinction, and $y$ as the lower detector QE/throughput results in high exposure times for relatively little additional scientific gain.  

\subsection{Exposure constraints}
\begin{footnotesize}
{\it Describe any constraints on the minimum or maximum exposure time per visit required (or alternatively, saturation limits).
Please comment on any constraints on the number of exposures in a visit.}
\end{footnotesize}\\

Exposure times for this DDF are primarily constrained by the need to reach limiting magnitudes $>$22\,mag in all passbands.  \wfirst will survey stars between $i\sim$19--25\,mag, so with \lsst we propose to monitor the brightest of the sample, $i\sim$19--23\,mag.  Please note that difference imaging techniques will enable us to push deeper than the nominal crowding limit. The required magnitude range fits well within \lsstns's dynamic range for 1$\times$30\,s or 2$\times$15\,s exposure.  Follow-up observations will be obtained from other ground-based resources for priority targets that are magnified to brightnesses which would saturate in \lsstns's images.  There is some benefit to 2$\times$15\,s exposure in that it allows for better cosmic ray rejection, and \lsst data on targets up to $\sim$0.75\,mag brighter.  

\subsection{Other constraints}
\begin{footnotesize}
{\it Any other constraints.}
\end{footnotesize}

We note strong synergies between this proposal and the proposal for a wide-area survey of the Galactic Plane (Street et al.), which will naturally include this DDF.  While the wide-area survey cannot substitute for the high-cadence, simultaneous coverage we propose here, it would obtain the lower-cadence, inter-season monitoring.  If the wide-area Plane survey goes ahead, {\it additional} inter-season observations specifically for the DDF would not be necessary.  

\subsection{Estimated time requirement}
\begin{footnotesize}
{\it Approximate total time requested for these observations, using the guidelines available at \url{https://github.com/lsst-pst/survey_strategy_wp}.}
\end{footnotesize}

The time required for the inter-season observations is 0.045\,hrs / day for $\sim$68\,days, giving a total of 3.06\,hrs per year.  

In addition, the time required for the simultaneous observations depends strongly on exactly when observations during the simultaneous viewing periods start and end.  The table below gives estimates of the time required per year for different examples.  The duration of each window assumes a \wfirst survey season duration of 72\,days.  Please note that, due to spacecraft constraints, it is very likely that the season dates will be those in the {\it first} column.  

\begin{table}[h!]
\centering
\begin{tabular}{llll}
\hline
Window              & Time [hrs/year] &   Window             &  Time [hrs/year] \\
\hline
Feb 12 -- Apr 24    &  24.07     &  Jan 15 -- Mar 31    & 13.32 \\
Aug 19 -- Oct 29    &  24.53     &  Aug 21 -- Oct 31    & 23.67\\
Total               &  48.6      &  Total               & 36.99\\
\hline
\end{tabular}
\end{table}

\vspace{.3in}

\begin{table}[ht]
    \centering
    \begin{tabular}{l|l|l|l}
        \toprule
        Properties & Importance \hspace{.3in} \\
        \midrule
        Image quality & 2    \\
        Sky brightness &  3 \\
        Individual image depth &  2 \\
        Co-added image depth &  3 \\
        Number of exposures in a visit   &  2 \\
        Number of visits (in a night)   & 2  \\ 
        Total number of visits &   2 \\
        Time between visits (in a night) &  1 \\
        Time between visits (between nights)  &   1 \\
        Long-term gaps between visits & 1 \\
        Other (please add other constraints as needed) & \\
        \bottomrule
    \end{tabular}
    \caption{{\bf Constraint Rankings:} Summary of the relative importance of various survey strategy constraints. Please rank the importance of each of these considerations, from 1=very important, 2=somewhat important, 3=not important. If a given constraint depends on other parameters in the table, but these other parameters are not important in themselves, please only mark the final constraint as important. For example, individual image depth depends on image quality, sky brightness, and number of exposures in a visit; if your science depends on the individual image depth but not directly on the other parameters, individual image depth would be `1' and the other parameters could be marked as `3', giving us the most flexibility when determining the composition of a visit, for example.}
        \label{tab:obs_constraints}
\end{table}

\subsection{Technical trades}
\begin{footnotesize}
{\it To aid in attempts to combine this proposed survey modification with others, please address the following questions:}
\begin{enumerate}
    \item {\it What is the effect of a trade-off between your requested survey footprint (area) and requested co-added depth or number of visits?}
    
    As this DDF proposal has a single pointing, the only possible trade-off is in terms of the frequency of visits.  Longer intervals between visits (in any filter) serves to increase the minimum microlensing event timescale to which the survey would be sensitive (meaning the minimum mass of the detectable lenses would increase), and the periods (in the case of variable targets).  It would provide lower cadence photometry of the anomalous lightcurve features that are critical to properly characterize the nature of binary lenses.  Planetary signatures would become increasingly difficult to distinguish, and the survey would be sensitive to stellar binaries.  
    
    \item {\it If not requesting a specific timing of visits, what is the effect of a trade-off between the uniformity of observations and the frequency of observations in time? e.g. a `rolling cadence' increases the frequency of visits during a short time period at the cost of fewer visits the rest of the time, making the overall sampling less uniform.}
    
    Both the simultaneous and inter-season observations require specific timing windows. 
    While a rolling cadence could provide better lightcurve coverage in some years, which would be advantageous, the poorer coverage in other years would serve to decrease the overall yield, since microlensing is a transient phenomenon, and new events are expected with each season.  Inconsistently-sampled lightcurves will make it more difficult to constrain the parallax (measured from the long-term skew in the lightcurve shape), and more likely that anomalous features could be missed or mis-characterized.
    
    Rolling cadence observations would be beneficial for the periodic variables types in our science case, however (UCDs, transits, RR~Lyrae in GCs, rise-times of CVs) since it would enable us to better constraint their optical periods.  Bearing in mind that higher-cadence \wfirst lightcurves will also be available, this is not a compelling case for rolling cadence; the main gain of this DDF is overlapping observations with \wfirst and \lsst for an extended baseline.  
    
    \item {\it What is the effect of a trade-off on the exposure time and number of visits (e.g. increasing the individual image depth but decreasing the overall number of visits)?}
    
    This would decrease the scientific yield of all of our science cases, which depend primarily on the time-sampling of the target lightcurves rather than the limiting magnitude of the imaging data.  
    
    \item {\it What is the effect of a trade-off between uniformity in number of visits and co-added depth? Is there any benefit to real-time exposure time optimization to obtain nearly constant single-visit limiting depth?}

    Our proposed science does not depend on the co-added image depth, so any trade-off with number of visits is likely to reduce the overall cadence and time-sampling, and hence to decrease the scientific yield.  There negative impacts of adjusting the exposure for consistent limiting depth are likely to outweigh the benefits in signal-to-noise, since this would inevitably come at the expense of cadence. 
    
    \item {\it Are there any other potential trade-offs to consider when attempting to balance this proposal with others which may have similar but slightly different requests?}
    
    The timing of the simultaneous windows can be moved earlier or later in the year {\bf (provided they still overlap with \wfirstns)}.  This results in the visibility of the field being more (towards mid-year) or less (towards the end of the year), and hence changes the number of visits per night.  If this is necessary, it would be best to neglect observations on those nights where visibility is shortest, since the minimum useful time per night is approximately the width of a typical event peak, which is set by the time required to cross the angular diameter of the source star ($\sim$hours).  That said, the microlensing parallax signal is larger at these times and easier to measure.  However, we note that shorter nightly observing windows would negatively impact all other science cases, so this would make the most sensible trade off.  
    
    The frequency and/or the timing of multi-filter exposures may be one source of trade-offs, for instance optimizing the filter selection to minimize overheads, or maintaining regular high cadence observations in $i$ and/or $z$ but reducing cadence in the bluer filters.
    
    The cadence of the inter-season gap observations could reduced (to a minimum of once every 3\,d) in order to maintain a high cadence during the simultaneous periods.    
    
\end{enumerate}
\end{footnotesize}
\pagebreak
\section{Performance Evaluation}
The most important criteria for evaluating observational strategies with respect to this proposal are the temporal and spatial overlap with the WFIRST Bulge Survey.  To this end, we propose the following metrics, for which hotlinks are given below (bold font) to public Github repositories.  \\

\noindent {\bf Survey Footprint Metrics}\\
We have developed a \href{https://github.com/LSST-TVSSC/software_tools/blob/master/SpacialOverlapMetric.py}{SpacialOverlapMetric} to evaluate the degree of spatial overlap between a given OpSim and a defined region, in this case the WFIRST footprint, which we need to be 100\%.  Further optimization of this DDF's footprint can be made by avoiding the nearby regions of high-extinction though the application of the existing \href{https://github.com/LSST-nonproject/sims\_maf\_contrib/tree/master/mafContrib/StarCounts/}{StarCounts} metric.\\

We have developed a tool to assist with optimizing the DDF footprint, \href{https://github.com/LSST-TVSSC/software_tools/blob/master/optimize_bulge_ddf_footprint.py} \\{\bf optimize\_bulge\_ddf\_footprint.py}, available from the TVS public Github repository. This incorporates a list of Bulge Globular Clusters compiled from \cite{minniti:2017}, for inclusion within the DDF pointing.  A minimum requirement on the number of GCs within the footprint is 2, but estimate that a typical number should be $\sim$9 clusters.\\

\noindent {\bf Temporal Overlap Metrics}\\
To ensure sufficient observations during the simultaneous viewing windows, we recommend maximizing the new \href{https://github.com/Somayeh91/sims_maf_contrib/blob/master/mafContrib/numObsInSurveyTimeOverlap.py}{\bf numObsInSurveyTimeOverlap} metric.  This computes the number of observations within specific time periods, which should be defined to match the WFIRST survey periods $\sim$Feb 12-- Apr 24 and $\sim$Aug 19 -- Oct 29, between $\sim$2026 to 2030, with the exact dates to be confirmed nearer mission launch. \\
Observations during the inter-season gaps can be evaluated by minimizing the \href{https://github.com/Somayeh91/sims_maf_contrib/blob/master/mafContrib/IntervalsBetweenObs.py}{\bf IntervalsBetweenObs} metric, which calculates the mean and median interval between observations within specific time periods.  The maximum scientifically useful median interval of observations would be 3\,days, beyond which insufficient data would be obtained for events of typical timescales ($\sim$20--30\,days) with which to constrain their parallax and lightcurve shape.  \\
We have also developed the \href{https://github.com/rachel3834/sims_maf_contrib/blob/master/mafContrib/CadenceOverVisibilityWindowMetric.py}{\bf CadenceOverVisbilityWindowMetric}, which is designed to compare the actual number of visits (in a given filter set) with the desired number of visits, calculated from the visibility window of the field for the given start and end dates, and desired cadence. The value of this metric, $M = [\sum_{\rm nights} (N_{\rm{visits,actual}} / N_{\rm{visits,desired}})]/N_{\rm{filters}}$, equals 1.0 if all visits are made.  For our purposes, the minimum useful duration of simultaneous observations per night is set by the delay in caustic crossing time between the two observatories.  The projected relative velocity, ${\overline{v}} = 1/(\pi_E t_E) \sim$500\,km/sec, giving delay of $\sim$50\,mins, sufficient time for two visits to this DDF.  To give a rough estimate of the minimum-useful metric value, a single-visit gives $M = [1 / 2]/2 = 0.25.$\\

\noindent{\bf Figures of Merit, $FoM$}\\
The metrics described above can be combined to define a $FoM = O(f) C(f),$ where $O(f)$ represents the SpacialOverlapMetric, $C(f)$ the cadence over visibility window metric, computed for the full six \wfirst observing seasons plus any pre-launch observations by \lsstns.  \\

\noindent{\bf Observational Criteria}\\
The Moon will cause unavoidable interruptions to our proposed cadence, since it passes through the Galactic Plane a number of times a year.  We note that the \lsst scheduler already contains a lunar avoidance metric, which should be sufficient to prefer fields at appropriate separations from the Moon during these periods.  We further suggest preferring $i$ and $z$ band observations only during these times.

\vspace{.6in}

\section{Special Data Processing}
\begin{footnotesize}
{\it Describe any data processing requirements beyond the standard LSST Data Management pipelines and how these will be achieved.}
\end{footnotesize}

The data acquired in the course of this DDF should meet the requirements of the LSST Data Management pipeline for automatic processing.  No special processing should be necessary.  

\section{Acknowledgements}
This work developed partly within the TVS Science Collaboration and the author acknowledge the support of TVS in the preparation of this paper. The authors acknowledge support from the Flatiron Institute and Heising-Simons Foundation for the development of this paper.

\section{}

\end{document}